\documentclass[%
 aip,
rsi,%
 amsmath,amssymb,
 reprint,%
]{revtex4-1}

\usepackage{graphicx}
\usepackage{dcolumn}
\usepackage{bm}
\begin{document}

\preprint{AIP/123-QED}

\title{Self-compression of stimulated Raman backscattering by flying focus}
\author{Z. Wu}
\author{Y. Zuo}%
\author{Z. Zhang}
\author{X. Wang}
\author{J. Mu}%
\author{X.D. Wang}
\author{B. Hu}
\author{J. Su}%
\author{Z. Li}
\author{X. Wei}%
\author{X. Zeng}
\email{zxm7311@sina.com}
\affiliation{Science and Technology on Plasma Physics Laboratory, Research Center of Laser Fusion, China Academic of Engineering Physics, Mianyang, Sichuan, China, 621900.}


\date{\today}

\begin{abstract}
  A novel regime of self-compression is proposed for plasma-based backward Raman amplification(BRA) upon flying focus. By using a pumping focus moving with a speed equal to the group velocity of stimulated Raman backscattering(SRBS), only a short part of SRBS which does always synchronize with the flying focus can be amplified. Due to the asymmetrical amplification, the pulse can be directly compressed in the linear stage of BRA. Therefore, instead of a short pulse, the Raman spontaneous or a long pulse can seed the BRA amplifiers. The regime is supported by the 2D particle-in-cell(PIC) simulation without a seed, presenting that the pump pulse is compressed from 26ps to 116fs, with an output amplitude comparable with the case of a well-synchronized short seed. This method provides a significant way to simplify the Raman amplifiers and overcome the issue of synchronization jitter between the pump and the seed.
\end{abstract}

\keywords{Plasma, flying focus, chromatic Raman compression, Backward Raman scattering }
\maketitle                            

Currently, laser powers of multiply petawatt has been reported by large laser facilities relying on chirped pulse amplification(CPA)\cite{Mourou85}. A further enhancement of laser power is impeded by the material damage of compression gratings. The difficult is proposed to be overcame by using plasma for laser amplification. The plasma amplifiers, based on backward Raman amplification (BRA)\cite{Malkin991,Malkin00,Mourou12,Ping04,Cheng05,Ren08} or strongly coupled stimulated Brillouin scattering(scSBS)\cite{Andreev06,Lancia13,Lancia16,Marques19} have shown the amplified intensities over $10^{16}~\rm{W/cm^2}$, indicating exawatt laser powers can be obtain within centimeter-diameter amplifiers.

 The BRA experiment has shown an amplified seed intensity far exceed that of the pump, however, the transfer efficiency(6.4\% for double passes, and 5.1\% for a single pass)\cite{Ren08,Turnbull12,Turnbull12s} and pump depletion (more than 70\% of the pump energy remained) were still low. One explanation attributes the low efficiency to the thermal effects from the pump before the seed arrived. Plasma heating can influence BRA by accelerating the electrons close to the phase velocity of plasma wave(about c/20). As a consequence, BRA would be suppressed by several negative effects such as particle trapping, Landau damping, and plasma wavebreaking. Another explanation is the idea condition for the BRA can not be fully satisfied by the experiment due to an extremely complex setup. For instance, for typical plasma amplifiers, several laser pulses are used in the experiment for the ionization pulses, pump and seed, respectively. The spatiotemporal overlap of them can hardly be kept due to the perturbation of the laser system.

Flying focus has been proposed to overcome the precursors and plasma heating in front of the seed\cite{Turbunll18,Dustin18}. In this scheme, a laser focus is able to move with an artificial velocity after a chromatically focusing system. By setting the focusing pump intensity just above the ionization threshold of the background gas, a ionization wave is generated with a moving velocity equal to the flying focus\cite{Turbunll18s}. For a flying velocity around $v=c$, the seed pulse can always be arranged just behind the ionization wave. Therefore, a clean environment, without pump heating and Raman spontaneous, is created in front of the seed. This scheme, however, requires a more precise synchronization of the pump and the seed, or the amplified seed will be extremely unstable. A stationary plasma-wave seed with a density structure of two-peaks is able to solve the issue of synchronization jitter\cite{Kenan17}. However, it is still a challenge job to produce such a plasma structure. To our knowledge, it has not be experimentally realized yet.
  \begin{figure}[htpb]
\includegraphics[width=3 in]{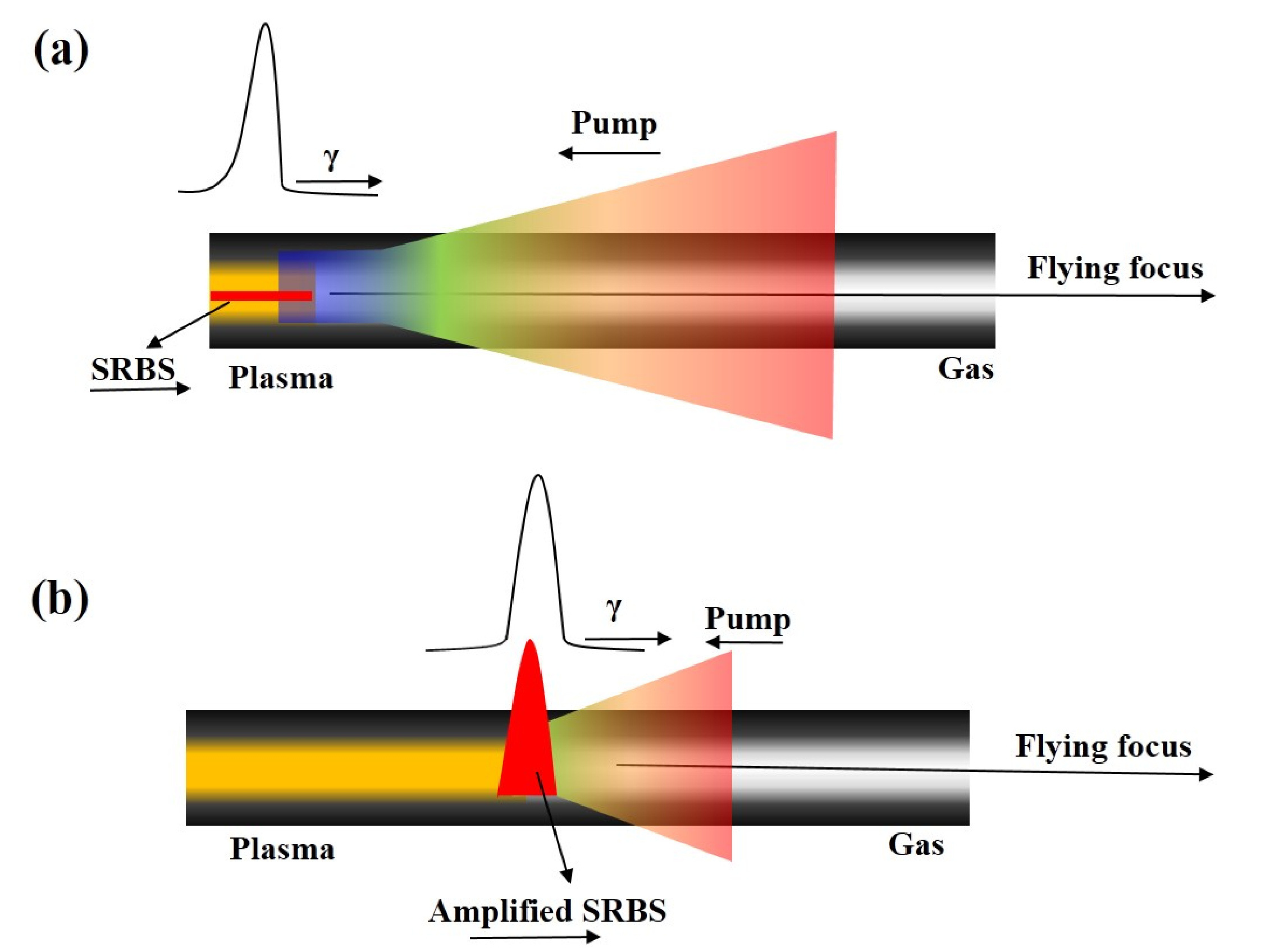}
\caption{\label{fig:shceme}Illustration of the self-compression regime due to SRBS by flying focus. The flying focus with a growth curve is set the same velocity(c) as SRBS but counter to the pump velocity(-c). (a)The plasma and SRBS is initially created at the entrance by flying focus. (b) The SRBS propagating with flying focus is amplified to a short pulse.}
\end{figure}

In this paper, we show for the first time the 2d particle-in-cell(PIC) simulation of BRA with flying focus. Moreover, it is proposed a self-compression regime to simplify the Raman amplifiers and solve the issue of the synchronization jitter. The scheme is displayed in Fig.\ref{fig:shceme}. As the focal intensity of the pump is slightly higher than the ionization threshold of the background gas, SRBS mainly occurs within the pump focus. By adjusting the flying velocity the same as the SRBS, the growth curve $\gamma$ would always overlap with a small part of initial SRBS, leading to an asymmetrical amplification. Consequently, the amplified pulse is compressed to around the width of $\gamma$ curve in the linear stage of BRA, as shown in Fig.\ref{fig:shceme}(b). A self-compression regime is generated to transfer the long disordered Raman noisy to a short coherent seed. Similarly, it can transfer a long seed to a short one, indicating the Raman spontaneous or a long pulse can seed Raman amplifiers.  As Raman spontaneous is a parasitic pulse of the pump, and a long duration seed can always partly overlap with the pump focus, the synchronization jitter can be naturally avoided.

In general, flying focus can be described by focusing a collimated laser beam with chirp and pulse group delay(PDG). The spatio-frequency E-field of the flying focus consists of a superposition of
Gaussian beams of different frequencies, with there waist located at different positions. Commonly, the spatio-temporal E-field can only be numerically obtained by the inverse Fourier transform. However, in the limit of large pulse chirp, it can be estimated analytically as\cite{Sainte}
 \begin{eqnarray}
\begin{array}{rcl}
a(z,t')=\frac{a_0(z,t'/\beta)}{\sqrt{1+z(t')^2}}e^{i\phi(z,t')}
\end{array}
\label{phase1}
\end{eqnarray}

With

\begin{eqnarray}
\begin{array}{rcl}
\phi(z,t')=arctan[z(t')]+\frac{kr^2}{2z(t')[1+z(t')^2]}+\frac{{t'}^2}{2\beta}
\end{array}
\label{phase2}
\end{eqnarray}

where $t'=t-z/c$, $z(t')=t'\tau/\beta-z/z_R$, $\beta$ is the pulse chirp, $\tau$ is the parameter of PGD, and $z_R$ is the Rayleigh length. The beam waist locates at the position of $z=t'\tau z_R/\beta$ for $z(t')=0$, indicating the focus spot is not static but variable along the propagating direction. By adjusting the pulse chirp $\beta$ and PGD $\tau$, the focus can move with a velocity from subluminal to superluminal\cite{Sainte,Dustin18,Dustin19}.

As the growth rate of the Langmuir wave and SRBS are the same in the linear stage, the initial Raman spontaneous $b_s$ can be estimated by the magnification of the energy intensity of Langmuir wave from thermal noisy to pump depletion\cite{Malkin14},
 \begin{eqnarray}
\begin{array}{rcl}
b_{s}\approx a_0\sqrt{\frac{W_{eT}}{W_{ek}}}=\frac{a_0e}{m_ec^2}\sqrt{\frac{T_e}{8\pi\varepsilon_0\lambda_a}}
\end{array}
\label{noisy}
\end{eqnarray}

where $T_e$ is the electron temperature, $e$ and $m_e$ is the electron charge and mass, and $\varepsilon_0$ is the permittivity of free space, and c is speed of light in vacuum, $W_{eT}\sim \frac{T_e\omega_p\omega_a}{4\pi\lambda_a}$ and $W_{ek}\sim\frac{I_0\omega_p}{\omega_a}$ are the energy density of Langmuir wave for the equilibrium thermal noise and pump depletion, respectively, for $I_0$ is the pump intensity, $\omega_p=\sqrt{\frac{n_ee^2}{m_e\varepsilon_0}}$ and $\omega_a$ are the frequency of the Langmuir wave and pump, and $n_e$ is the electron plasma density.

\begin{figure}[htpb]
\includegraphics[width=1.5 in]{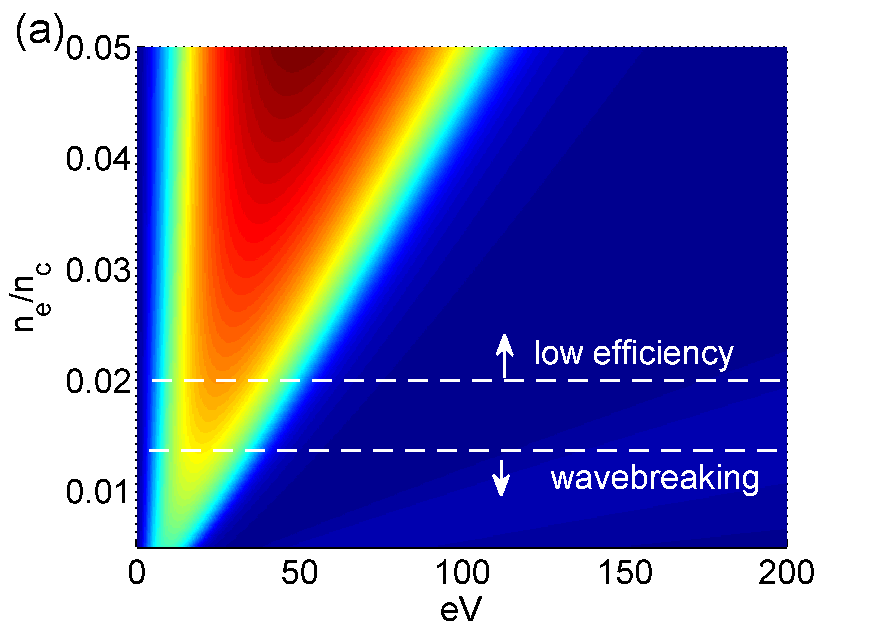}
\includegraphics[width=1.5 in]{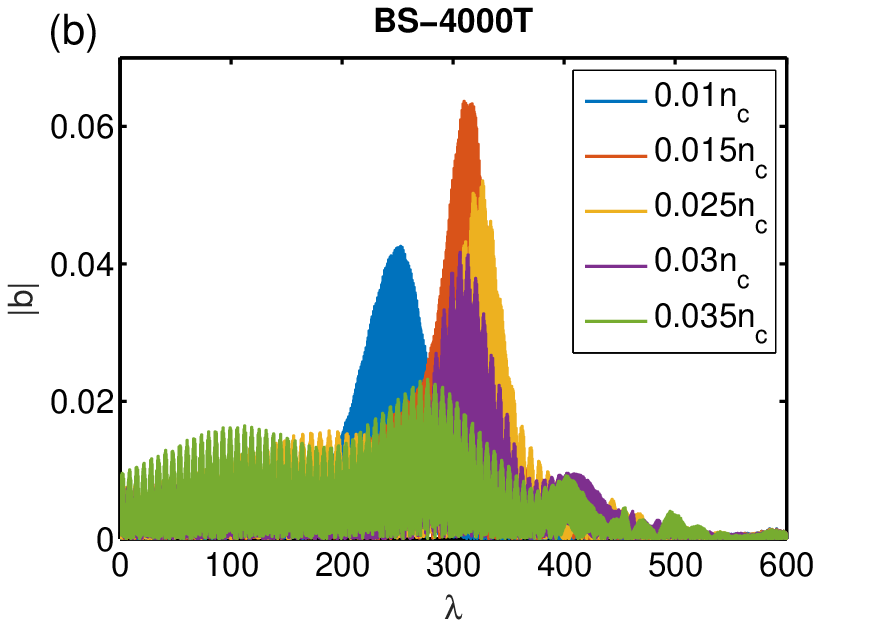}\\
\includegraphics[width=1.5 in]{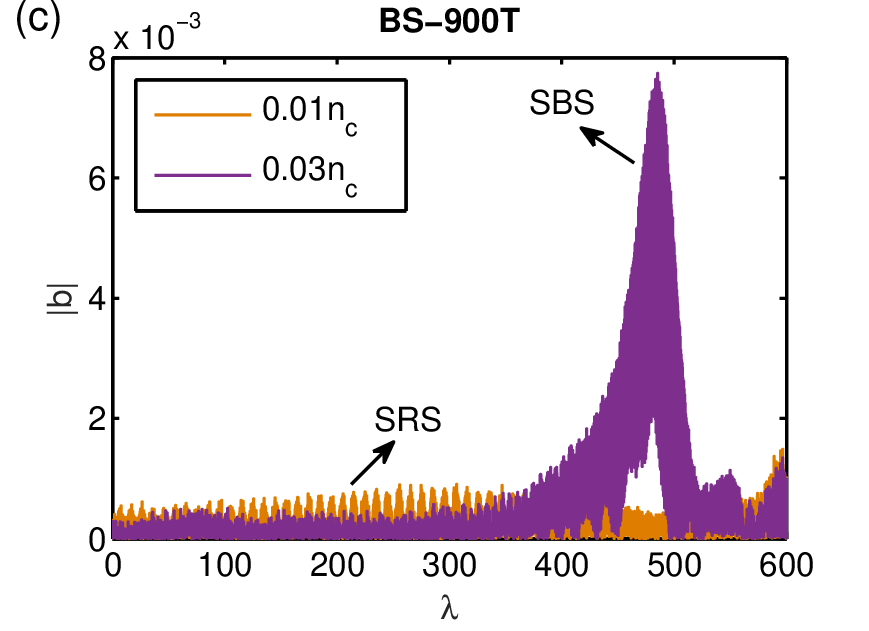}
\includegraphics[width=1.5 in]{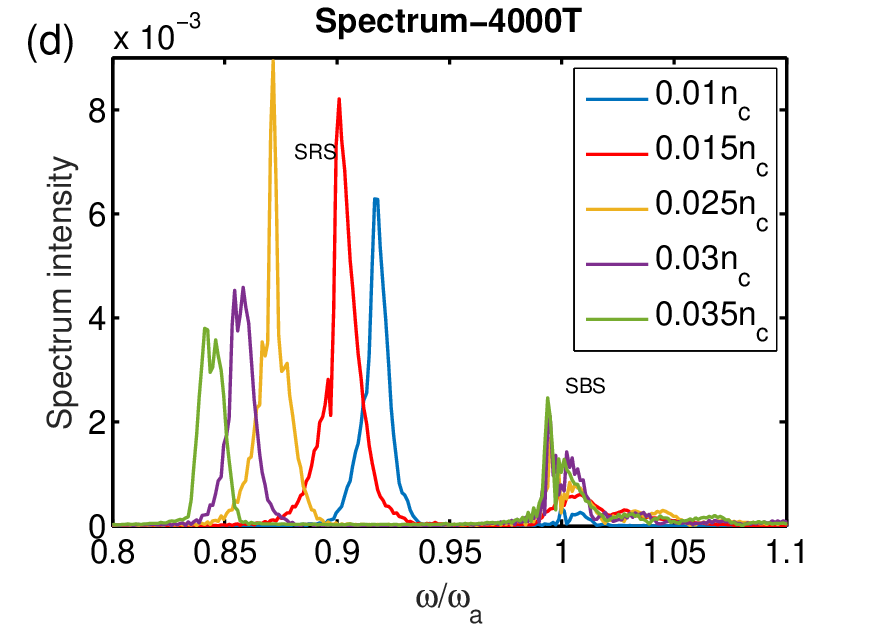}
\caption{\label{fig:growthrate}(a)Numerical solution of SRBS growth rate with various plasma densities and temperatures. (b)On-axial $|b|$ from spontaneous at different plasma densities. (c)Spectrum of on-axial $|b|$ from spontaneous at different plasma densities. (d)On-axial $|b|$ at t=900T for plasma density of 0.015$n_c$ and 0.03$n_c$. In the simulation, $|b|$ is the absolute amplitude of SRBS, $T_e=25$eV, $a_0=0.01$, $n_c=\omega^2_am_e\varepsilon_0/e^2$ is the critical plasma density.}
\end{figure}

An optimal window of parameter is expected for the self-compression regime. The pump intensity has little space to be improved because it has to be slightly above the ionization threshold(0.01 in our simulation). Therefore, the BRA performance is mainly optimized from plasma density and temperature. The idea linear SRBS growth rate is $\gamma=\sqrt{\omega_p\omega_a}/2$. With damping and frequency detuning, it changes to $Re(\lambda)$ for a langmuir wave $f\propto e^{\lambda t}$. $\lambda$ satisfies\cite{Balakin20}
 \begin{eqnarray}
\begin{array}{rcl}
\lambda^2+(\nu+i\delta\omega)\lambda-\gamma^2=0,
\end{array}
\label{growthrate}
\end{eqnarray}
where $\nu=\nu_{ei}/4+\nu_{Ld}$ is the total damping, for collision damping $\nu_{ei}=2.9\times10^6Zn_e[cm^{-3}]\Lambda[eV]^{-3/2}$, where $\Lambda$ is the Coulomb logarithm, $Z$ is the plasma ion
charge\cite{Balakin11,Clark03,Johnson17,Malkin09}, Landua damping $\nu_{Ld}=\omega_p\sqrt{\pi}/{{(2q_T)}^{3/2}}exp(-1/2q_T-3/2)$, $\delta\omega=k_fT_e/\omega_p m_e$ is the frequency detuning due to thermal chirp, where $q_T=k_fT_e/\omega_p m_e$, $k_f$ is the wave number of the Langmuir wave\cite{Balakin11,Malkin09}. The numerical solution of SRBS growth rate is shown in Fig.\ref{fig:growthrate}, illustrating an optimal temperature can be found for each plasma density.

 In order to further optimize the plasma density, we used a 2D particle-in-cell(PIC) code OPIC2D\cite{zhang12,Zhang14,zhang17} to simulate the growth of Raman spontaneous with different plasma densities. The best amplification is obtained at density of 0.15$n_c$, as shown in Fig.\ref{fig:growthrate}(b), corresponding to an optimal plasma temperature around 25eV. The major limitation of plasma density is from scSBS effect. As the pump is also used for ionization pulse, the ionization wave in front of SRBS can reflect pump to form an scSBS seed. Although the reflectivity is low, the intensity accumulating with the ionization wave can achieve to the same level as the pump, as shown in the on-axis SRBS at the density of 0.03$n_c$[Fig.\ref{fig:growthrate}(c)]. Despite of a low pump intensity(the amplitude is 0.01, corresponding to an intensity of $2.7\times10^{14}~\rm{W/cm^2}$ for the wavelength of 1 $\mu$m), the condition of scSBS can still be satisfied because of a low plasma temperature(25 eV), according to the threshold of scSBS $I>1.1\times10^{13}T_e^{3/2}n_c/n\sqrt{1-n/n_c}\lambda~\rm{W/cm^2}$. The scSBS effect becomes significant as the plasma density increases over 0.025nc. Typically, the interfringes generated by SRBS and scSBS appear in the amplified pulse, and SBS takes up an increasing proportion in the spectrum, as shown in Fig.\ref{fig:growthrate}(b)and(c).

\begin{figure}[htpb]
\includegraphics[width=1.5 in]{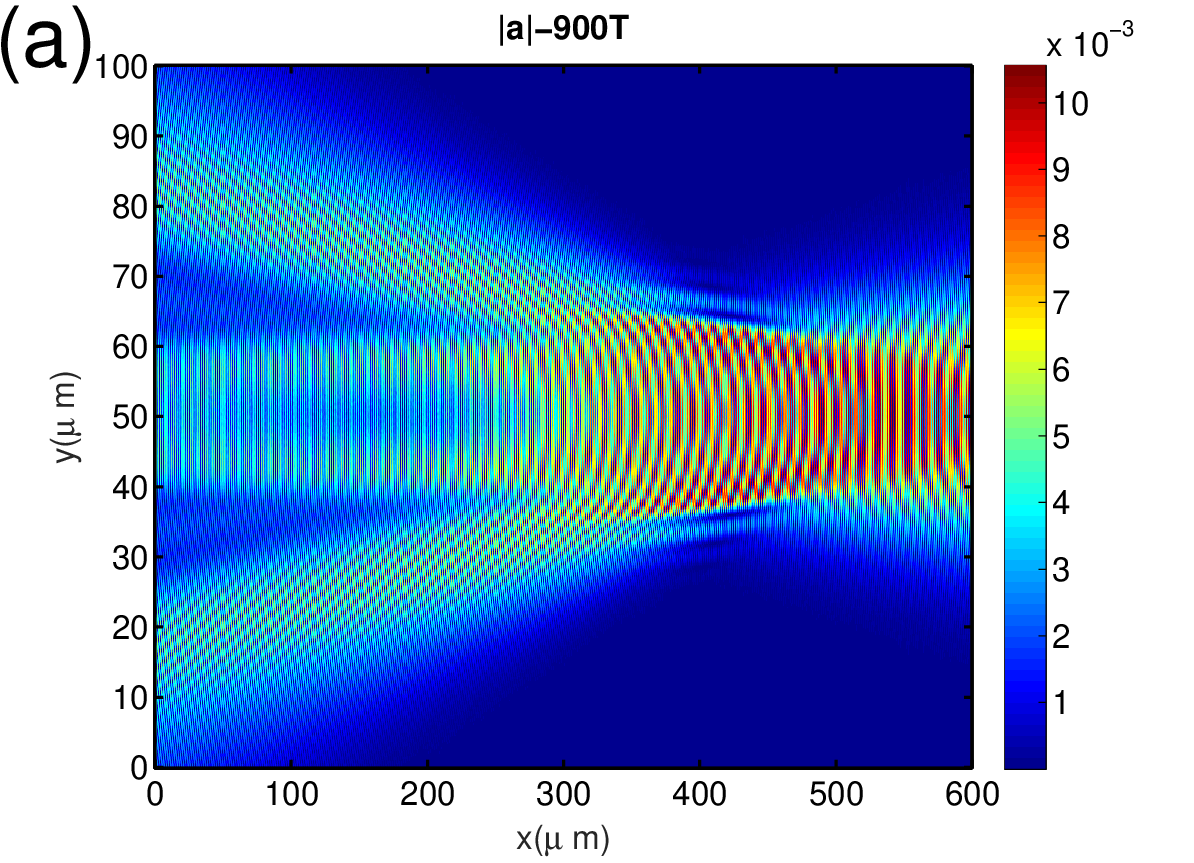}
\includegraphics[width=1.5 in]{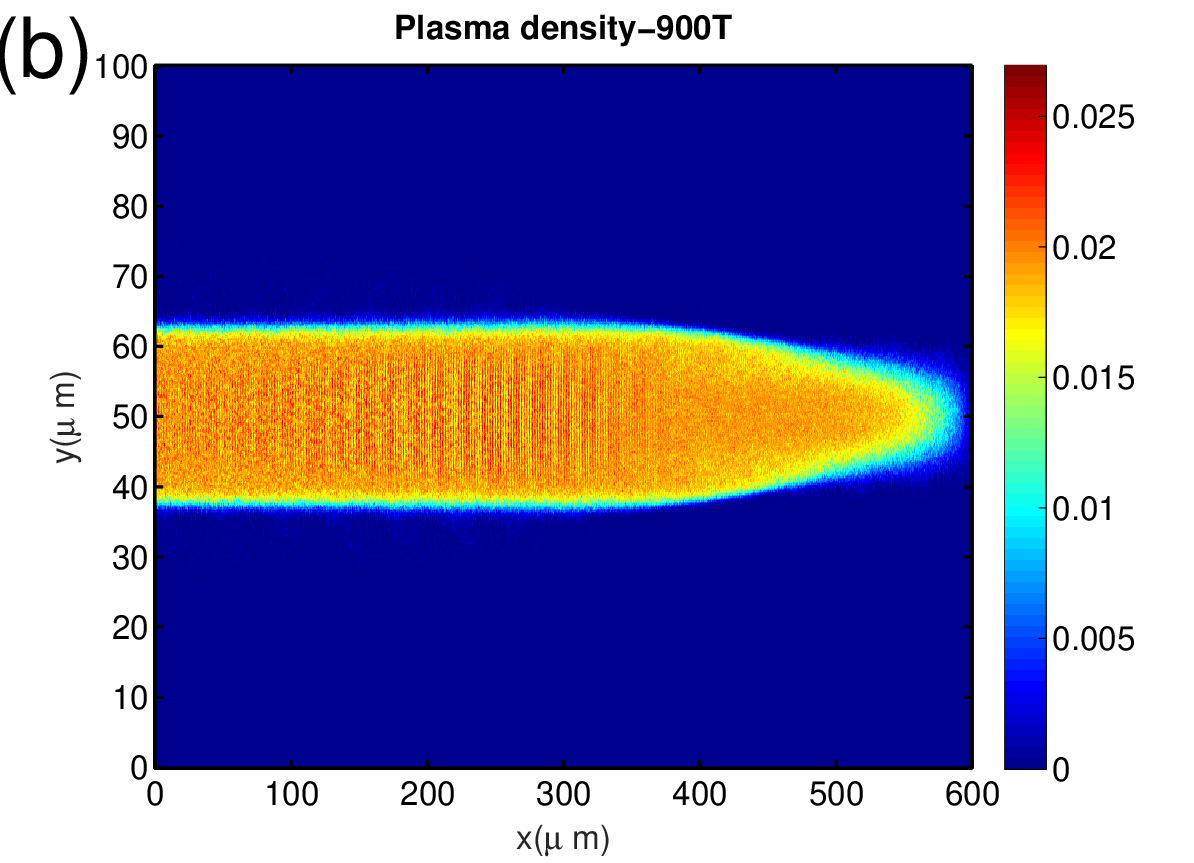}\\
\includegraphics[width=3 in]{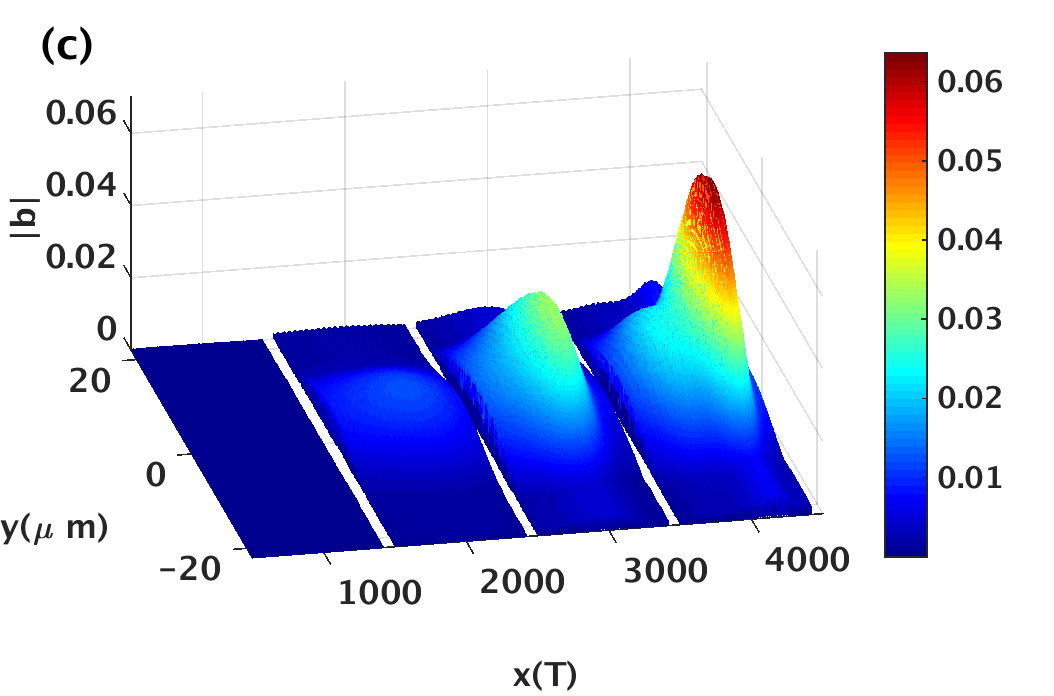}
\caption{\label{fig:2dresult}(a)Flying focus at t=900T, where T=3.3fs is the single cycle of 1 $\mu$m laser, $|a|$ is absolute valvule of normalized pump amplitude. The velocity of flying focus and moving window is the same(0.99c),
 so the focus is static in the simulation window. (b)Plasma density at t=900T, the gas at right boundary of the moving window is kept neutral. (c)2D PIC simulation result of SRBS amplitude $|b|$ developed from Raman spontaneous for $n_e=0.015n_c$, $T_e=25~\rm{eV}$.}
\end{figure}

 In the 2D simulation, the ionization and BRA process with flying focus are simulated in a moving window of $600\lambda \times100\lambda$ for the z-y plane, with cells of $\Delta_z=0.1\lambda$ and $\Delta_y=0.2\lambda$, and $16\times2$ particles for each cell, where $z$ and $y$ are the prorogating and transverse direction, respectively. A 4-mm-long and 50-$\mu$m-wide hydrogen is applied for the background gas. For the interaction length of 4 mm, the pump duration is about 26 ps, with a central wavelength of 1 $\mu$m. The initial pump focus has a waist of 5 $\mu$m, with a maximum amplitude of 0.02. After it countering propagates along the z axial of the moving window, the beam waist is magnified to about 10 $\mu$m due to ionization-induced defocusing, so the peak amplitude is reduced to 0.01, which is slightly above the ionization threshold of Hydrogen, as shown in Fig.\ref{fig:2dresult}(a). Beyond the focus about 200 $\mu$m, the background gas at the right boundary of the moving window is kept neutral, as shown in the plasma density[Fig.\ref{fig:2dresult}(b)] obtained by an Ammosov-Delone-Krainov tunnel ionization model(ADK).

The optimal 2d simulation result of self-compression process is shown in Fig\ref{fig:2dresult}(c). The initial spontaneous amplitude of $\sim3\times10^{-5}~\rm{W/cm^2}$ is obtained by inverse fast fourier transform(IFFT) of the wavelength from 1100 nm-1200 nm. After 2 mm amplification, the pump starts to be depleted as the interaction reaches the nonlinear stage. The pulse amplitude is further amplified to about 0.066 after 4-mm amplification. Due to the diffractive effect, the amplified SRBS beam is diverged from 10 $\mu m$ to 20 $\mu m$. The total transfer efficiency is about 40\%. The simulation result indicates a short laser pulse with an amplitude far exceed that of the pump can be obtained from Raman spontaneous upon flying focus. As only one pump pulse is applied for the amplification, the Raman amplifiers can be significantly simplified.

\begin{figure}[htpb]
\includegraphics[width=1.5 in]{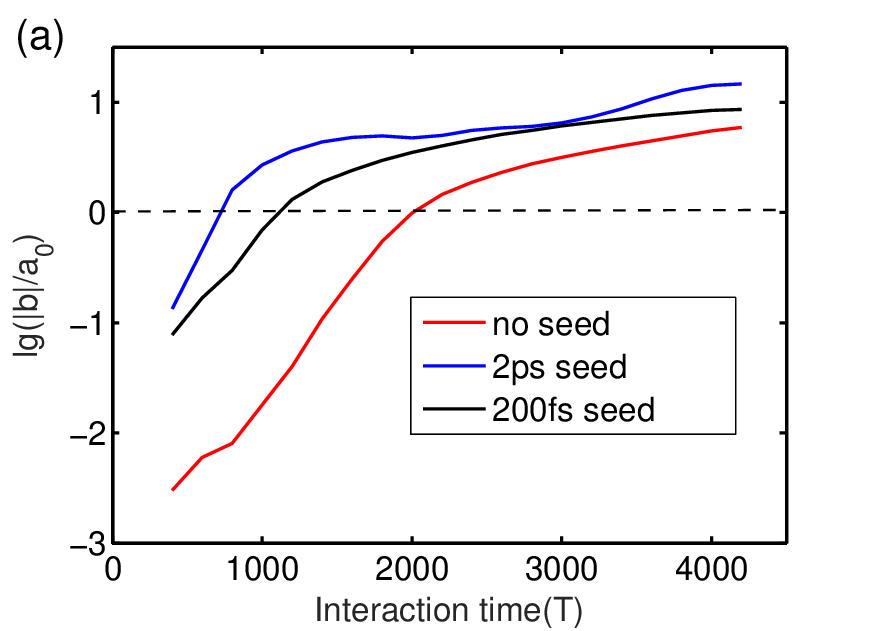}
\includegraphics[width=1.5 in]{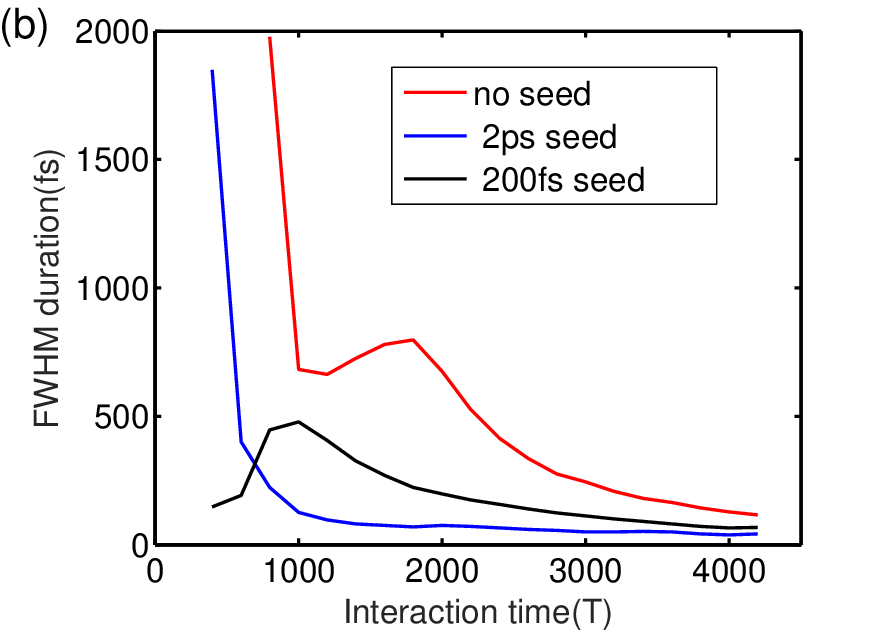}\\
\includegraphics[width=1.5 in]{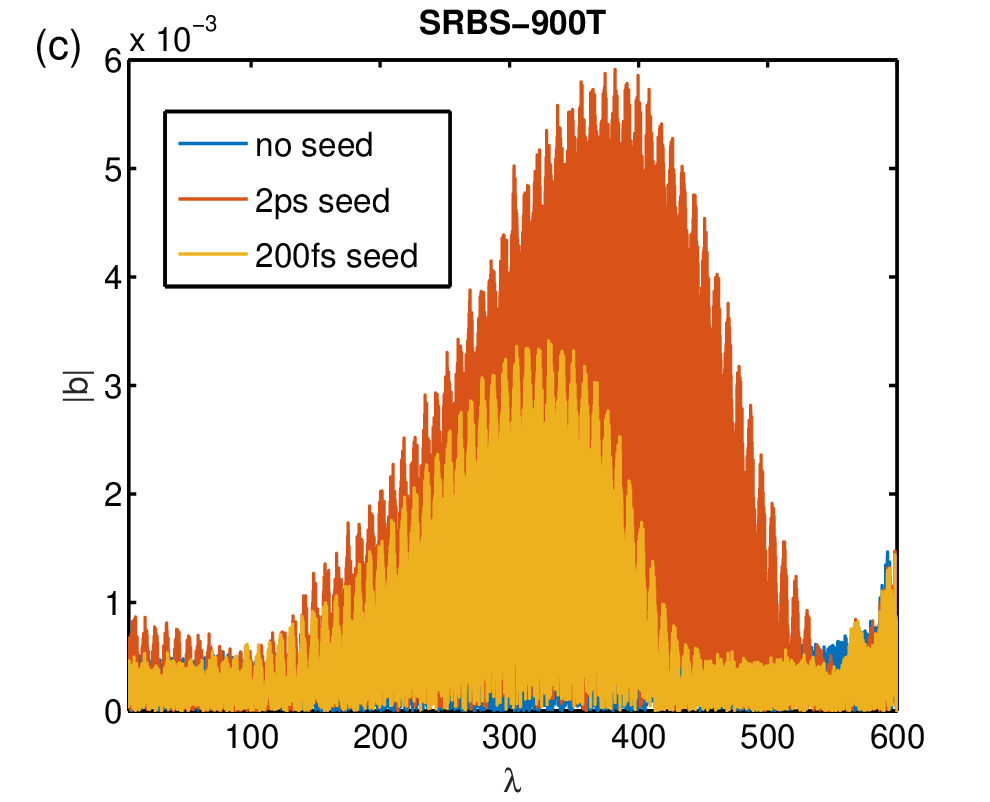}
\includegraphics[width=1.5 in]{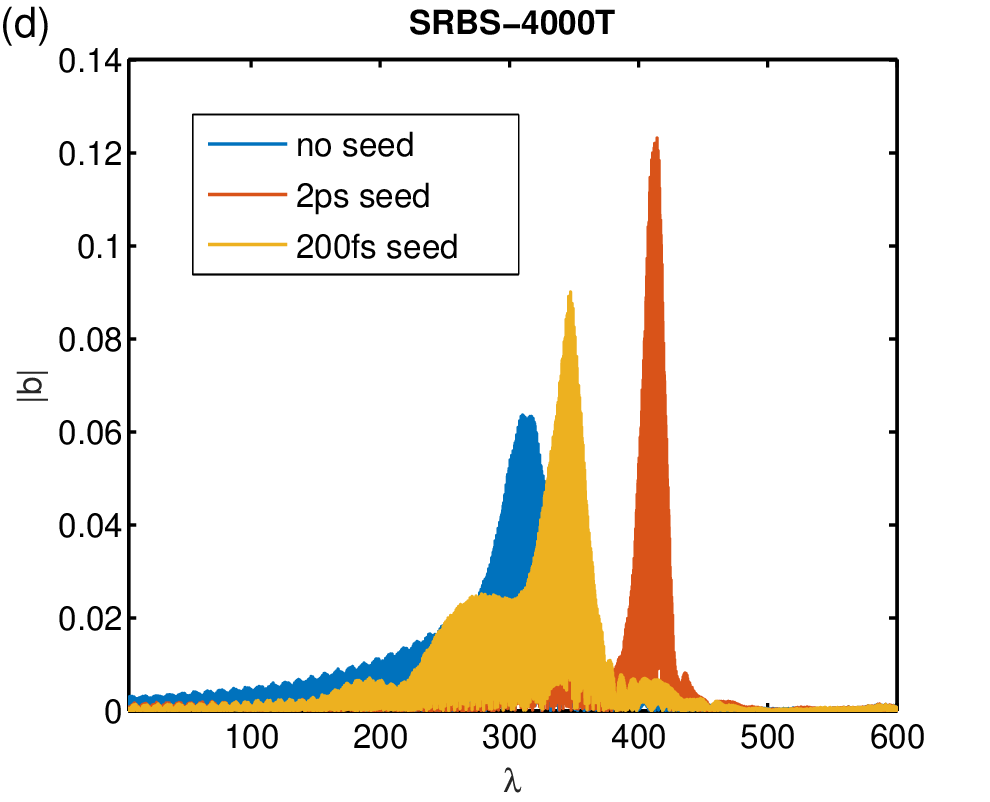}
\caption{\label{fig:comparison}Simulation result of (a) peak intensity and (b) FWHM duration with a 2 ps seed, a 200 fs seed, and Raman spontaneous.(c) On-axial pulse shapes of SRBS at t=900T. (d) On-axial pulse shapes of SRBS at t=4000T.}
\end{figure}

The simulation results with a well-synchronized short seed and a long-duration seed are displayed in Fig.\ref{fig:comparison}. Both the long and short seeds have a central wavelength 1140 nm, an amplitude of 0.001, a beam waist of 100 $\mu$m(for a Rayleigh length more than 4 mm). The long seed with a duration of 2 ps covers the full moving window while the short seed with a duration of 200 fs locates at 400$\mu$m of the moving window.  As shown in Fig.\ref{fig:comparison}(a), although the initial amplitude of the Raman spontaneous is 1/30 as the seeds, the amplified amplitudes tend to be comparable after the 4-mm amplification. This is mainly due to the growth rate $\gamma$ degrades from exponential to near linear after the pump is depleted. Despite a strong seed enable the interaction to reach the nonlinear stage earlier, the growth then is approached by the case of a Raman noisy seed. The evolution of pulse durations are displayed in Fig.\ref{fig:comparison}(b): both the Raman spontaneous and 2ps seed are directly compressed in the linear stage, despite the duration of Raman spontaneous slightly returns back during a short time. The 200fs seed is stretched in the linear stage because it is already narrower than the growth rate curve. In the nonlinear, all pulses are further compressed to 116 fs, 65 fs, and 40 fs by the Raman spontaneous, the 200 fs seed and the 20
ps seed, respectively.

The best result is obtained by the long-duration seed, with an output peak amplitude and a transfer efficiency of 0.12 and 28.3\% respectively, which are higher than 0.09 and 24.4\% obtained by the short seed. This is mainly due to a better envelope match of the SRBS maximum with gain maximum. The concept of envelope match has been proposed in the seed-ionizing BRA\cite{Zhang14}. Here we show a similar regime by using a long-duration seed and flying focus. In the linear stage, the SRBS maximum has a subluminous speed\cite{Malkin991}, a long-duration seed can accelerate the maximum moving by providing a plasma wave to the following interaction from the pulse front. Therefore, the linear amplified seed peak is more close to the pump focus(400$\lambda$), as shown in Fig.\ref{fig:comparison}(c). Conversely, in the nonlinear stage when the SRBS peak has a superluminous speed\cite{Malkin991}, the acceleration of the SRBS maximum moving becomes a negative factor. However, this issue can be overcame here because the SRBS can not exceed the ionization wave producing by the flying focus. As shown in Fig.\ref{fig:comparison}(d), the amplified peak of the 2 ps seed still does not exceed the pump focus at 4000T. Therefore, different from the previous conclusion that a short seed with a sharp front is preferred for BRA, a long-duration seed here achieves to an even better result.

One thing to be noted is that the numerical noisy in the PIC code is several orders of magnitude as the thermal fluctuation in plasma. The numerical noisy can be reduced by increasing particles per cell, however, it brings huge calculation for the simulation. Alternatively, as the amplification of thermal fluctuation to the numerical noisy is very close to the idea linear amplification, an extra interaction length can be estimated as $L=ln(\frac{b_n}{b_s})\frac{c}{\gamma}$, where $b_s$ and $b_n$ are the SRBS due to thermal fluctuation and numerical noisy, respectively. For $a_0=0.01$, $n_e=0.02n_c$, $T_e=5$ eV, $\lambda_a=1~\mu$m, it has $b_{s}\approx10^{-9}\approx b_n/4000$ according to Eq.\ref{noisy}, which agrees well with the result obtained by $\delta-$correct noise\cite{Balakin20}. Hence, the extra interaction length is estimated bout $L=0.85$ mm. The transfer efficiency will be reduced to about 16\% because of this extra interaction length. However, as Raman spontaneous can almost perfectly match the resonant frequency and spatio-temporal synchronization with the pump pulse, probably it can lead to an even higher transfer efficiency in the experiment. For instance, the highest efficiency reported by plasma-based Raman amplifiers so far is from Raman spontaneous\cite{Vieux17}.

The amplification of Raman spontaneous may be suppressed by the frequency and velocity detuning between the flying focus and SRBS. Frequency detuning is usually produced by plasma inhomogeneity, pulse chirp and plasma heating. Here, we mainly investigate the influence from the plasma inhomogeneity as the pulse chirp can be compensated by a well-designed plasma gradient, and plasma heating can be avoided by flying focus\cite{Turbunll18}. With frequency detuning, the linear growth rate becomes $\gamma'=\sqrt{\gamma^2-\delta\omega^2/4}$ according to Eq.\ref{growthrate}, indicating ${\delta\omega}$ can not exceed a threshold of $2\gamma$ for SRBS growth. However, in a nonuniform plasma, the SRBS can not be fully broken down even for a density fluctuation far exceed the threshold because there is always part of plasma satisfies $\delta\omega<2\gamma$. To investigate the influence from plasma inhomogeneity, a random density fluctuation from $10\%-30\%$ was used in the PIC model. As shown in Tab.\ref{tab:detuning}, the amplification still keep well at the density fluctuation of $10\%$.  The amplified $|b|$ is more sensitive to the velocity of flying focus, as shown in Tab.\ref{tab:detuning}. Especially for the velocity above $c$, the amplified amplitude decreases to 0.038 at a velocity of 1.02c. In the experiment, the precisely control of the focus velocity can be realized by adjusting the distance of 2 compression gratings\cite{Dustin18}.

 \begin{table}[htpb]
\centering
\caption{\label{tab:detuning} Amplified peak amplitudes with various plasma density fluctuation and flying velocities, for $a_0=0.01$, $T_e=25$eV, $n=0.015 n_c$.}
\begin{tabular}{llllll}
\hline
\itshape Density fluctuation($\%$) &0 & 5 &10 &20&30\\
\itshape Amplified amplitude($|b|$) &0.066 & 0.064 & 0.058 & 0.050&0.047\\
\itshape Flying velocity(c) &0.98 & 0.99 & 1& 1.02&1.05\\
\itshape Amplified amplitude($|b|$) &0.053& 0.066 & 0.066 & 0.038&0.027\\
\hline
\end{tabular}
\end{table}

Another issue of flying focus is the limitation of interaction aperture. In order to migrate precursors and plasma heating, it requires producing an ionization wave a static short distance in front of SRBS\cite{Turbunll18}. Commonly, the Rayleigh length of the pump focus has to be short, or the intensity would be sufficiently strong to generate plasma far ahead of the beam waist. For instance of a Gaussian beam, $z_R=w_0^2/\lambda\approx6.4$ mm for $w_0=100~\rm{\mu}$m,$\lambda=1~\rm{\mu}$m, where $w_0$ is the beam waist,$\lambda$ is the wavelength, implying the pump will start to produce plasma 6.4 mm ahead of the flying focus. As a result, both the precursors and plasma heating can not be migrated. Also, it can not be used for self-compression. To solve the problem, a random phase plate can be employed before the achromatical lens to magnify the beam waist without changing the Rayleigh length. The random phase plate is widely used in large laser facilities for a beam waist more than 200 $\mu m$\cite{Ping091}. To further enlarge the beam focus to millimeter or centimeter, a microns array can be used to merge a series small focus into a large one.

In summary, we propose a new self-compression regime for plasma-based Raman amplifiers. The use of flying focus can generate a growth rate moving with the same velocity as SRBS, so only a short part of SRBS always synchronizing with the flying focus can be amplified, leading to the self-compression in the linear stage of BRA. This regime is supported by a 2D PIC simulation without a seed, showing a transfer efficiency around 18\% after 4 mm interaction. Meanwhile, a long-duration seed is demonstrated possible for BRA in this regime. It leads to an even a better result than a short seed by improving the envelope match of the SRBS maximum with gain maximum. Above of all, a new way to generate short laser pulses by Raman scattering is presented. Moveover, by reducing the complexity and avoiding the synchronization jitter of plasma-based Raman amplifiers, it is a promising method to generate ultrahigh laser power.

This work was partly supported by National Key Program for S and T Research and Development (Grant No. 2018YFA0404804), the National Natural Science Foundation of China(Grant No. 11875240) and the Science and Technology on Plasma Physics Laboratory Fund (Grant No. 6142A0403010417).

\bibliography{ref}

\providecommand{\noopsort}[1]{}\providecommand{\singleletter}[1]{#1}%
\begin{thebibliography}{30}%
\makeatletter
\providecommand \@ifxundefined [1]{%
 \@ifx{#1\undefined}
}%
\providecommand \@ifnum [1]{%
 \ifnum #1\expandafter \@firstoftwo
 \else \expandafter \@secondoftwo
 \fi
}%
\providecommand \@ifx [1]{%
 \ifx #1\expandafter \@firstoftwo
 \else \expandafter \@secondoftwo
 \fi
}%
\providecommand \natexlab [1]{#1}%
\providecommand \enquote  [1]{``#1''}%
\providecommand \bibnamefont  [1]{#1}%
\providecommand \bibfnamefont [1]{#1}%
\providecommand \citenamefont [1]{#1}%
\providecommand \href@noop [0]{\@secondoftwo}%
\providecommand \href [0]{\begingroup \@sanitize@url \@href}%
\providecommand \@href[1]{\@@startlink{#1}\@@href}%
\providecommand \@@href[1]{\endgroup#1\@@endlink}%
\providecommand \@sanitize@url [0]{\catcode `\\12\catcode `\$12\catcode
  `\&12\catcode `\#12\catcode `\^12\catcode `\_12\catcode `\%12\relax}%
\providecommand \@@startlink[1]{}%
\providecommand \@@endlink[0]{}%
\providecommand \url  [0]{\begingroup\@sanitize@url \@url }%
\providecommand \@url [1]{\endgroup\@href {#1}{\urlprefix }}%
\providecommand \urlprefix  [0]{URL }%
\providecommand \Eprint [0]{\href }%
\providecommand \doibase [0]{http://dx.doi.org/}%
\providecommand \selectlanguage [0]{\@gobble}%
\providecommand \bibinfo  [0]{\@secondoftwo}%
\providecommand \bibfield  [0]{\@secondoftwo}%
\providecommand \translation [1]{[#1]}%
\providecommand \BibitemOpen [0]{}%
\providecommand \bibitemStop [0]{}%
\providecommand \bibitemNoStop [0]{.\EOS\space}%
\providecommand \EOS [0]{\spacefactor3000\relax}%
\providecommand \BibitemShut  [1]{\csname bibitem#1\endcsname}%
\let\auto@bib@innerbib\@empty
\bibitem [{\citenamefont {Strickland}\ and\ \citenamefont
  {Mourou}(1985)}]{Mourou85}%
  \BibitemOpen
  \bibfield  {author} {\bibinfo {author} {\bibfnamefont {D.}~\bibnamefont
  {Strickland}}\ and\ \bibinfo {author} {\bibfnamefont {G.}~\bibnamefont
  {Mourou}},\ }\href@noop {} {\bibfield  {journal} {\bibinfo  {journal} {Opt.
  Commun.}\ }\textbf {\bibinfo {volume} {56}},\ \bibinfo {pages} {219}
  (\bibinfo {year} {1985})}\BibitemShut {NoStop}%
\bibitem [{\citenamefont {Malkin}, \citenamefont {Shvets},\ and\ \citenamefont
  {Fisch}(1999)}]{Malkin991}%
  \BibitemOpen
  \bibfield  {author} {\bibinfo {author} {\bibfnamefont {V.~M.}\ \bibnamefont
  {Malkin}}, \bibinfo {author} {\bibfnamefont {G.}~\bibnamefont {Shvets}}, \
  and\ \bibinfo {author} {\bibfnamefont {N.~J.}\ \bibnamefont {Fisch}},\
  }\href@noop {} {\bibfield  {journal} {\bibinfo  {journal} {Phys. Rev. Lett}\
  }\textbf {\bibinfo {volume} {82}},\ \bibinfo {pages} {4448} (\bibinfo {year}
  {1999})}\BibitemShut {NoStop}%
\bibitem [{\citenamefont {Malkin}, \citenamefont {Shvets},\ and\ \citenamefont
  {Fisch}(2000)}]{Malkin00}%
  \BibitemOpen
  \bibfield  {author} {\bibinfo {author} {\bibfnamefont {V.~M.}\ \bibnamefont
  {Malkin}}, \bibinfo {author} {\bibfnamefont {G.}~\bibnamefont {Shvets}}, \
  and\ \bibinfo {author} {\bibfnamefont {N.~J.}\ \bibnamefont {Fisch}},\
  }\href@noop {} {\bibfield  {journal} {\bibinfo  {journal} {Phys. Rev. Lett}\
  }\textbf {\bibinfo {volume} {84}},\ \bibinfo {pages} {1208} (\bibinfo {year}
  {2000})}\BibitemShut {NoStop}%
\bibitem [{\citenamefont {Mourou}\ \emph {et~al.}(2012)\citenamefont {Mourou},
  \citenamefont {Fisch}, \citenamefont {Malkin}, \citenamefont {Toroker},
  \citenamefont {Khazanov}, \citenamefont {Sergeev}, \citenamefont {Tajima},\
  and\ \citenamefont {Garrec}}]{Mourou12}%
  \BibitemOpen
  \bibfield  {author} {\bibinfo {author} {\bibfnamefont {G.~A.}\ \bibnamefont
  {Mourou}}, \bibinfo {author} {\bibfnamefont {N.~J.}\ \bibnamefont {Fisch}},
  \bibinfo {author} {\bibfnamefont {V.~M.}\ \bibnamefont {Malkin}}, \bibinfo
  {author} {\bibfnamefont {Z.}~\bibnamefont {Toroker}}, \bibinfo {author}
  {\bibfnamefont {E.~A.}\ \bibnamefont {Khazanov}}, \bibinfo {author}
  {\bibfnamefont {A.~M.}\ \bibnamefont {Sergeev}}, \bibinfo {author}
  {\bibfnamefont {T.}~\bibnamefont {Tajima}}, \ and\ \bibinfo {author}
  {\bibfnamefont {B.~L.}\ \bibnamefont {Garrec}},\ }\href@noop {} {\bibfield
  {journal} {\bibinfo  {journal} {Optics Communications}\ }\textbf {\bibinfo
  {volume} {285}},\ \bibinfo {pages} {720} (\bibinfo {year}
  {2012})}\BibitemShut {NoStop}%
\bibitem [{\citenamefont {Ping}, \citenamefont {Cheng},\ and\ \citenamefont
  {Suckewer}(2004)}]{Ping04}%
  \BibitemOpen
  \bibfield  {author} {\bibinfo {author} {\bibfnamefont {Y.}~\bibnamefont
  {Ping}}, \bibinfo {author} {\bibfnamefont {W.}~\bibnamefont {Cheng}}, \ and\
  \bibinfo {author} {\bibfnamefont {S.}~\bibnamefont {Suckewer}},\ }\href@noop
  {} {\bibfield  {journal} {\bibinfo  {journal} {Phys. Rev. Lett}\ }\textbf
  {\bibinfo {volume} {92}},\ \bibinfo {pages} {175007} (\bibinfo {year}
  {2004})}\BibitemShut {NoStop}%
\bibitem [{\citenamefont {Cheng}\ \emph {et~al.}(2005)\citenamefont {Cheng},
  \citenamefont {Avitzour}, \citenamefont {Ping},\ and\ \citenamefont
  {Suckewer}}]{Cheng05}%
  \BibitemOpen
  \bibfield  {author} {\bibinfo {author} {\bibfnamefont {W.}~\bibnamefont
  {Cheng}}, \bibinfo {author} {\bibfnamefont {Y.}~\bibnamefont {Avitzour}},
  \bibinfo {author} {\bibfnamefont {Y.}~\bibnamefont {Ping}}, \ and\ \bibinfo
  {author} {\bibfnamefont {S.}~\bibnamefont {Suckewer}},\ }\href@noop {}
  {\bibfield  {journal} {\bibinfo  {journal} {Phys. Rev. Lett}\ }\textbf
  {\bibinfo {volume} {94}},\ \bibinfo {pages} {045003} (\bibinfo {year}
  {2005})}\BibitemShut {NoStop}%
\bibitem [{\citenamefont {Ren}\ \emph {et~al.}(2008)\citenamefont {Ren},
  \citenamefont {Li}, \citenamefont {Morozov}, \citenamefont {Suckewer},
  \citenamefont {Yampolsky}, \citenamefont {Malkin},\ and\ \citenamefont
  {Fisch}}]{Ren08}%
  \BibitemOpen
  \bibfield  {author} {\bibinfo {author} {\bibfnamefont {J.}~\bibnamefont
  {Ren}}, \bibinfo {author} {\bibfnamefont {S.}~\bibnamefont {Li}}, \bibinfo
  {author} {\bibfnamefont {A.}~\bibnamefont {Morozov}}, \bibinfo {author}
  {\bibfnamefont {S.}~\bibnamefont {Suckewer}}, \bibinfo {author}
  {\bibfnamefont {N.~A.}\ \bibnamefont {Yampolsky}}, \bibinfo {author}
  {\bibfnamefont {V.~M.}\ \bibnamefont {Malkin}}, \ and\ \bibinfo {author}
  {\bibfnamefont {N.~J.}\ \bibnamefont {Fisch}},\ }\href@noop {} {\bibfield
  {journal} {\bibinfo  {journal} {Phys. Plasmas}\ }\textbf {\bibinfo {volume}
  {15}},\ \bibinfo {pages} {056702} (\bibinfo {year} {2008})}\BibitemShut
  {NoStop}%
\bibitem [{\citenamefont {Andreev}\ \emph {et~al.}(2006)\citenamefont
  {Andreev}, \citenamefont {Riconda}, \citenamefont {Tikhonchuk},\ and\
  \citenamefont {Weber}}]{Andreev06}%
  \BibitemOpen
  \bibfield  {author} {\bibinfo {author} {\bibfnamefont {A.~A.}\ \bibnamefont
  {Andreev}}, \bibinfo {author} {\bibfnamefont {C.}~\bibnamefont {Riconda}},
  \bibinfo {author} {\bibfnamefont {V.~T.}\ \bibnamefont {Tikhonchuk}}, \ and\
  \bibinfo {author} {\bibfnamefont {S.}~\bibnamefont {Weber}},\ }\href@noop {}
  {\bibfield  {journal} {\bibinfo  {journal} {Phys. Plasma}\ }\textbf {\bibinfo
  {volume} {13}},\ \bibinfo {pages} {053110} (\bibinfo {year}
  {2006})}\BibitemShut {NoStop}%
\bibitem [{\citenamefont {Lancia}\ \emph {et~al.}(2010)\citenamefont {Lancia},
  \citenamefont {Marqu\`{e}s}, \citenamefont {Nakatsutsumi}, \citenamefont
  {Riconda}, \citenamefont {Weber}, \citenamefont {H\"{u}ller}, \citenamefont
  {A.Man\v{c}i\'{c}}, \citenamefont {Antici}, \citenamefont {Tikhonchuk},
  \citenamefont {H\'{e}ron}, \citenamefont {Audebert},\ and\ \citenamefont
  {Fuchs}}]{Lancia13}%
  \BibitemOpen
  \bibfield  {author} {\bibinfo {author} {\bibfnamefont {L.}~\bibnamefont
  {Lancia}}, \bibinfo {author} {\bibfnamefont {J.~R.}\ \bibnamefont
  {Marqu\`{e}s}}, \bibinfo {author} {\bibfnamefont {M.}~\bibnamefont
  {Nakatsutsumi}}, \bibinfo {author} {\bibfnamefont {C.}~\bibnamefont
  {Riconda}}, \bibinfo {author} {\bibfnamefont {S.}~\bibnamefont {Weber}},
  \bibinfo {author} {\bibfnamefont {S.}~\bibnamefont {H\"{u}ller}}, \bibinfo
  {author} {\bibnamefont {A.Man\v{c}i\'{c}}}, \bibinfo {author} {\bibfnamefont
  {P.}~\bibnamefont {Antici}}, \bibinfo {author} {\bibfnamefont {V.~T.}\
  \bibnamefont {Tikhonchuk}}, \bibinfo {author} {\bibfnamefont
  {A.}~\bibnamefont {H\'{e}ron}}, \bibinfo {author} {\bibfnamefont
  {P.}~\bibnamefont {Audebert}}, \ and\ \bibinfo {author} {\bibfnamefont
  {J.}~\bibnamefont {Fuchs}},\ }\href@noop {} {\bibfield  {journal} {\bibinfo
  {journal} {Phys. Rev. Lett.}\ }\textbf {\bibinfo {volume} {104}},\ \bibinfo
  {pages} {025001} (\bibinfo {year} {2010})}\BibitemShut {NoStop}%
\bibitem [{\citenamefont {Lancia}\ \emph {et~al.}(2016)\citenamefont {Lancia},
  \citenamefont {Giribono}, \citenamefont {Vassura}, \citenamefont
  {Chiaramello}, \citenamefont {Riconda}, \citenamefont {Weber}, \citenamefont
  {Castan}, \citenamefont {Chatelain}, \citenamefont {Frank}, \citenamefont
  {Gangolf}, \citenamefont {Quinn}, \citenamefont {Fuchs},\ and\ \citenamefont
  {Marqu\`{e}s}}]{Lancia16}%
  \BibitemOpen
  \bibfield  {author} {\bibinfo {author} {\bibfnamefont {L.}~\bibnamefont
  {Lancia}}, \bibinfo {author} {\bibfnamefont {A.}~\bibnamefont {Giribono}},
  \bibinfo {author} {\bibfnamefont {L.}~\bibnamefont {Vassura}}, \bibinfo
  {author} {\bibfnamefont {M.}~\bibnamefont {Chiaramello}}, \bibinfo {author}
  {\bibfnamefont {C.}~\bibnamefont {Riconda}}, \bibinfo {author} {\bibfnamefont
  {S.}~\bibnamefont {Weber}}, \bibinfo {author} {\bibfnamefont
  {A.}~\bibnamefont {Castan}}, \bibinfo {author} {\bibfnamefont
  {A.}~\bibnamefont {Chatelain}}, \bibinfo {author} {\bibfnamefont
  {A.}~\bibnamefont {Frank}}, \bibinfo {author} {\bibfnamefont
  {T.}~\bibnamefont {Gangolf}}, \bibinfo {author} {\bibfnamefont {M.~N.}\
  \bibnamefont {Quinn}}, \bibinfo {author} {\bibfnamefont {J.}~\bibnamefont
  {Fuchs}}, \ and\ \bibinfo {author} {\bibfnamefont {J.-R.}\ \bibnamefont
  {Marqu\`{e}s}},\ }\href@noop {} {\bibfield  {journal} {\bibinfo  {journal}
  {Phys. Rev. Lett.}\ }\textbf {\bibinfo {volume} {116}},\ \bibinfo {pages}
  {075001} (\bibinfo {year} {2016})}\BibitemShut {NoStop}%
\bibitem [{\citenamefont {Marqu\'{e}s}\ \emph {et~al.}(2019)\citenamefont
  {Marqu\'{e}s}, \citenamefont {Lancia}, \citenamefont {Gangolf}, \citenamefont
  {Blecher}, \citenamefont {{n}os}, \citenamefont {Fuchs}, \citenamefont
  {Willi}, \citenamefont {Amiranoff}, \citenamefont {Berger}, \citenamefont
  {Chiaramello}, \citenamefont {Weber},\ and\ \citenamefont
  {Riconda}}]{Marques19}%
  \BibitemOpen
  \bibfield  {author} {\bibinfo {author} {\bibfnamefont {J.-R.}\ \bibnamefont
  {Marqu\'{e}s}}, \bibinfo {author} {\bibfnamefont {L.}~\bibnamefont {Lancia}},
  \bibinfo {author} {\bibfnamefont {T.}~\bibnamefont {Gangolf}}, \bibinfo
  {author} {\bibfnamefont {M.}~\bibnamefont {Blecher}}, \bibinfo {author}
  {\bibfnamefont {S.~B.}\ \bibnamefont {{n}os}}, \bibinfo {author}
  {\bibfnamefont {J.}~\bibnamefont {Fuchs}}, \bibinfo {author} {\bibfnamefont
  {O.}~\bibnamefont {Willi}}, \bibinfo {author} {\bibfnamefont
  {F.}~\bibnamefont {Amiranoff}}, \bibinfo {author} {\bibfnamefont {R.~L.}\
  \bibnamefont {Berger}}, \bibinfo {author} {\bibfnamefont {M.}~\bibnamefont
  {Chiaramello}}, \bibinfo {author} {\bibfnamefont {S.}~\bibnamefont {Weber}},
  \ and\ \bibinfo {author} {\bibfnamefont {C.}~\bibnamefont {Riconda}},\
  }\href@noop {} {\bibfield  {journal} {\bibinfo  {journal} {PHYSICAL REVIEW
  X}\ }\textbf {\bibinfo {volume} {9}},\ \bibinfo {pages} {021008} (\bibinfo
  {year} {2019})}\BibitemShut {NoStop}%
\bibitem [{\citenamefont {Turnbull}\ \emph
  {et~al.}(2012{\natexlab{a}})\citenamefont {Turnbull}, \citenamefont {Li},
  \citenamefont {Morozov},\ and\ \citenamefont {S.Suckewer}}]{Turnbull12}%
  \BibitemOpen
  \bibfield  {author} {\bibinfo {author} {\bibfnamefont {D.}~\bibnamefont
  {Turnbull}}, \bibinfo {author} {\bibfnamefont {S.}~\bibnamefont {Li}},
  \bibinfo {author} {\bibfnamefont {A.}~\bibnamefont {Morozov}}, \ and\
  \bibinfo {author} {\bibnamefont {S.Suckewer}},\ }\href@noop {} {\bibfield
  {journal} {\bibinfo  {journal} {Phys. Plasmas}\ }\textbf {\bibinfo {volume}
  {19}},\ \bibinfo {pages} {073103} (\bibinfo {year}
  {2012}{\natexlab{a}})}\BibitemShut {NoStop}%
\bibitem [{\citenamefont {Turnbull}\ \emph
  {et~al.}(2012{\natexlab{b}})\citenamefont {Turnbull}, \citenamefont {Li},
  \citenamefont {Morozov},\ and\ \citenamefont {S.Suckewer}}]{Turnbull12s}%
  \BibitemOpen
  \bibfield  {author} {\bibinfo {author} {\bibfnamefont {D.}~\bibnamefont
  {Turnbull}}, \bibinfo {author} {\bibfnamefont {S.}~\bibnamefont {Li}},
  \bibinfo {author} {\bibfnamefont {A.}~\bibnamefont {Morozov}}, \ and\
  \bibinfo {author} {\bibnamefont {S.Suckewer}},\ }\href@noop {} {\bibfield
  {journal} {\bibinfo  {journal} {Phys. Plasmas}\ }\textbf {\bibinfo {volume}
  {19}},\ \bibinfo {pages} {083109} (\bibinfo {year}
  {2012}{\natexlab{b}})}\BibitemShut {NoStop}%
\bibitem [{\citenamefont {Turnbull}\ \emph
  {et~al.}(2018{\natexlab{a}})\citenamefont {Turnbull}, \citenamefont {Bucht},
  \citenamefont {Davies}, \citenamefont {Haberberger}, \citenamefont {Kessler},
  \citenamefont {Shaw},\ and\ \citenamefont {Froula1}}]{Turbunll18}%
  \BibitemOpen
  \bibfield  {author} {\bibinfo {author} {\bibfnamefont {D.}~\bibnamefont
  {Turnbull}}, \bibinfo {author} {\bibfnamefont {S.}~\bibnamefont {Bucht}},
  \bibinfo {author} {\bibfnamefont {A.}~\bibnamefont {Davies}}, \bibinfo
  {author} {\bibfnamefont {D.}~\bibnamefont {Haberberger}}, \bibinfo {author}
  {\bibfnamefont {T.}~\bibnamefont {Kessler}}, \bibinfo {author} {\bibfnamefont
  {J.~L.}\ \bibnamefont {Shaw}}, \ and\ \bibinfo {author} {\bibfnamefont
  {D.~H.}\ \bibnamefont {Froula1}},\ }\href@noop {} {\bibfield  {journal}
  {\bibinfo  {journal} {Phys. Rev. Lett.}\ }\textbf {\bibinfo {volume} {120}},\
  \bibinfo {pages} {024801} (\bibinfo {year} {2018}{\natexlab{a}})}\BibitemShut
  {NoStop}%
\bibitem [{\citenamefont {Froula}\ \emph {et~al.}(2018)\citenamefont {Froula},
  \citenamefont {Turnbull}, \citenamefont {Davies}, \citenamefont {Kessler},
  \citenamefont {Haberberger}, \citenamefont {Palastro}, \citenamefont {Bahk},
  \citenamefont {Begishev}, \citenamefont {Boni}, \citenamefont {Bucht},
  \citenamefont {Katz1},\ and\ \citenamefont {Shaw}}]{Dustin18}%
  \BibitemOpen
  \bibfield  {author} {\bibinfo {author} {\bibfnamefont {D.~H.}\ \bibnamefont
  {Froula}}, \bibinfo {author} {\bibfnamefont {D.}~\bibnamefont {Turnbull}},
  \bibinfo {author} {\bibfnamefont {A.~S.}\ \bibnamefont {Davies}}, \bibinfo
  {author} {\bibfnamefont {T.~J.}\ \bibnamefont {Kessler}}, \bibinfo {author}
  {\bibfnamefont {D.}~\bibnamefont {Haberberger}}, \bibinfo {author}
  {\bibfnamefont {J.~P.}\ \bibnamefont {Palastro}}, \bibinfo {author}
  {\bibfnamefont {S.-W.}\ \bibnamefont {Bahk}}, \bibinfo {author}
  {\bibfnamefont {I.~A.}\ \bibnamefont {Begishev}}, \bibinfo {author}
  {\bibfnamefont {R.}~\bibnamefont {Boni}}, \bibinfo {author} {\bibfnamefont
  {S.}~\bibnamefont {Bucht}}, \bibinfo {author} {\bibfnamefont
  {J.}~\bibnamefont {Katz1}}, \ and\ \bibinfo {author} {\bibfnamefont {J.~L.}\
  \bibnamefont {Shaw}},\ }\href@noop {} {\bibfield  {journal} {\bibinfo
  {journal} {Nature photonics}\ }\textbf {\bibinfo {volume} {12}},\ \bibinfo
  {pages} {262} (\bibinfo {year} {2018})}\BibitemShut {NoStop}%
\bibitem [{\citenamefont {Turnbull}\ \emph
  {et~al.}(2018{\natexlab{b}})\citenamefont {Turnbull}, \citenamefont {Franke},
  \citenamefont {Katz}, \citenamefont {Palastro}, \citenamefont {Begishev},
  \citenamefont {Boni}, \citenamefont {Bromage}, \citenamefont {Milder},
  \citenamefont {Shaw},\ and\ \citenamefont {Froula}}]{Turbunll18s}%
  \BibitemOpen
  \bibfield  {author} {\bibinfo {author} {\bibfnamefont {D.}~\bibnamefont
  {Turnbull}}, \bibinfo {author} {\bibnamefont {Franke}}, \bibinfo {author}
  {\bibfnamefont {J.}~\bibnamefont {Katz}}, \bibinfo {author} {\bibfnamefont
  {J.~P.}\ \bibnamefont {Palastro}}, \bibinfo {author} {\bibfnamefont {I.~A.}\
  \bibnamefont {Begishev}}, \bibinfo {author} {\bibfnamefont {R.}~\bibnamefont
  {Boni}}, \bibinfo {author} {\bibfnamefont {J.}~\bibnamefont {Bromage}},
  \bibinfo {author} {\bibfnamefont {A.~L.}\ \bibnamefont {Milder}}, \bibinfo
  {author} {\bibfnamefont {J.~L.}\ \bibnamefont {Shaw}}, \ and\ \bibinfo
  {author} {\bibfnamefont {D.~H.}\ \bibnamefont {Froula}},\ }\href@noop {}
  {\bibfield  {journal} {\bibinfo  {journal} {Phys. Rev. Lett.}\ }\textbf
  {\bibinfo {volume} {120}},\ \bibinfo {pages} {225001} (\bibinfo {year}
  {2018}{\natexlab{b}})}\BibitemShut {NoStop}%
\bibitem [{\citenamefont {Kenan~Qu}\ and\ \citenamefont
  {Fisch}(2017)}]{Kenan17}%
  \BibitemOpen
  \bibfield  {author} {\bibinfo {author} {\bibfnamefont {I.~B.}\ \bibnamefont
  {Kenan~Qu}}\ and\ \bibinfo {author} {\bibfnamefont {N.~J.}\ \bibnamefont
  {Fisch}},\ }\href@noop {} {\bibfield  {journal} {\bibinfo  {journal} {Phys.
  Rev. Lett.}\ }\textbf {\bibinfo {volume} {118}},\ \bibinfo {pages} {164801}
  (\bibinfo {year} {2017})}\BibitemShut {NoStop}%
\bibitem [{\citenamefont {SAINTE-MARIE}, \citenamefont {GOBERT},\ and\
  \citenamefont {QU\'{E}R\'{E}}(2017)}]{Sainte}%
  \BibitemOpen
  \bibfield  {author} {\bibinfo {author} {\bibfnamefont {A.}~\bibnamefont
  {SAINTE-MARIE}}, \bibinfo {author} {\bibfnamefont {O.}~\bibnamefont
  {GOBERT}}, \ and\ \bibinfo {author} {\bibfnamefont {F.}~\bibnamefont
  {QU\'{E}R\'{E}}},\ }\href@noop {} {\bibfield  {journal} {\bibinfo  {journal}
  {Optica}\ }\textbf {\bibinfo {volume} {4}},\ \bibinfo {pages} {1298}
  (\bibinfo {year} {2017})}\BibitemShut {NoStop}%
\bibitem [{\citenamefont {Froula}\ \emph {et~al.}(2019)\citenamefont {Froula},
  \citenamefont {Palastro}, \citenamefont {Turnbull}, \citenamefont {Davies},
  \citenamefont {Nguyen}, \citenamefont {Howard}, \citenamefont {Ramsey},
  \citenamefont {Franke}, \citenamefont {Bahk}, \citenamefont {Begishev},
  \citenamefont {Boni}, \citenamefont {Bromage}, \citenamefont {Bucht},
  \citenamefont {Follett}, \citenamefont {Haberberger}, \citenamefont
  {Jenkins}, \citenamefont {Katz}, \citenamefont {Kessler}, \citenamefont
  {Shaw},\ and\ \citenamefont {Vieira}}]{Dustin19}%
  \BibitemOpen
  \bibfield  {author} {\bibinfo {author} {\bibfnamefont {D.~H.}\ \bibnamefont
  {Froula}}, \bibinfo {author} {\bibfnamefont {J.~P.}\ \bibnamefont
  {Palastro}}, \bibinfo {author} {\bibfnamefont {D.}~\bibnamefont {Turnbull}},
  \bibinfo {author} {\bibfnamefont {A.}~\bibnamefont {Davies}}, \bibinfo
  {author} {\bibfnamefont {L.}~\bibnamefont {Nguyen}}, \bibinfo {author}
  {\bibfnamefont {A.}~\bibnamefont {Howard}}, \bibinfo {author} {\bibfnamefont
  {D.}~\bibnamefont {Ramsey}}, \bibinfo {author} {\bibfnamefont
  {P.}~\bibnamefont {Franke}}, \bibinfo {author} {\bibfnamefont {S.-W.}\
  \bibnamefont {Bahk}}, \bibinfo {author} {\bibfnamefont {I.~A.}\ \bibnamefont
  {Begishev}}, \bibinfo {author} {\bibfnamefont {R.}~\bibnamefont {Boni}},
  \bibinfo {author} {\bibfnamefont {J.}~\bibnamefont {Bromage}}, \bibinfo
  {author} {\bibfnamefont {S.}~\bibnamefont {Bucht}}, \bibinfo {author}
  {\bibfnamefont {R.~K.}\ \bibnamefont {Follett}}, \bibinfo {author}
  {\bibfnamefont {D.}~\bibnamefont {Haberberger}}, \bibinfo {author}
  {\bibfnamefont {G.~W.}\ \bibnamefont {Jenkins}}, \bibinfo {author}
  {\bibfnamefont {J.}~\bibnamefont {Katz}}, \bibinfo {author} {\bibfnamefont
  {T.~J.}\ \bibnamefont {Kessler}}, \bibinfo {author} {\bibfnamefont {J.~L.}\
  \bibnamefont {Shaw}}, \ and\ \bibinfo {author} {\bibfnamefont
  {J.}~\bibnamefont {Vieira}},\ }\href@noop {} {\bibfield  {journal} {\bibinfo
  {journal} {Phys. Plasmas}\ }\textbf {\bibinfo {volume} {26}},\ \bibinfo
  {pages} {032109} (\bibinfo {year} {2019})}\BibitemShut {NoStop}%
\bibitem [{\citenamefont {Malkin}\ and\ \citenamefont
  {Fisch}(2014)}]{Malkin14}%
  \BibitemOpen
  \bibfield  {author} {\bibinfo {author} {\bibfnamefont {V.~M.}\ \bibnamefont
  {Malkin}}\ and\ \bibinfo {author} {\bibfnamefont {N.~J.}\ \bibnamefont
  {Fisch}},\ }\href@noop {} {\bibfield  {journal} {\bibinfo  {journal} {Eur.
  Phys. J. Special Topics}\ }\textbf {\bibinfo {volume} {223}},\ \bibinfo
  {pages} {1157} (\bibinfo {year} {2014})}\BibitemShut {NoStop}%
\bibitem [{\citenamefont {Balakin}, \citenamefont {Levin},\ and\ \citenamefont
  {Skobelev}(2020)}]{Balakin20}%
  \BibitemOpen
  \bibfield  {author} {\bibinfo {author} {\bibfnamefont {A.~A.}\ \bibnamefont
  {Balakin}}, \bibinfo {author} {\bibfnamefont {D.~S.}\ \bibnamefont {Levin}},
  \ and\ \bibinfo {author} {\bibfnamefont {S.~A.}\ \bibnamefont {Skobelev}},\
  }\href@noop {} {\bibfield  {journal} {\bibinfo  {journal} {Phys. Rev. A}\
  }\textbf {\bibinfo {volume} {102}},\ \bibinfo {pages} {013516} (\bibinfo
  {year} {2020})}\BibitemShut {NoStop}%
\bibitem [{\citenamefont {Balakin}\ \emph {et~al.}(2011)\citenamefont
  {Balakin}, \citenamefont {Fisch}, \citenamefont {Fraiman}, \citenamefont
  {Malkin},\ and\ \citenamefont {Toroker}}]{Balakin11}%
  \BibitemOpen
  \bibfield  {author} {\bibinfo {author} {\bibfnamefont {A.~A.}\ \bibnamefont
  {Balakin}}, \bibinfo {author} {\bibfnamefont {N.~J.}\ \bibnamefont {Fisch}},
  \bibinfo {author} {\bibfnamefont {G.~M.}\ \bibnamefont {Fraiman}}, \bibinfo
  {author} {\bibfnamefont {V.~M.}\ \bibnamefont {Malkin}}, \ and\ \bibinfo
  {author} {\bibfnamefont {Z.}~\bibnamefont {Toroker}},\ }\href@noop {}
  {\bibfield  {journal} {\bibinfo  {journal} {Phys. Plasmas}\ }\textbf
  {\bibinfo {volume} {18}},\ \bibinfo {pages} {102311} (\bibinfo {year}
  {2011})}\BibitemShut {NoStop}%
\bibitem [{\citenamefont {Clark}\ and\ \citenamefont {Fisch}(2003)}]{Clark03}%
  \BibitemOpen
  \bibfield  {author} {\bibinfo {author} {\bibfnamefont {D.~S.}\ \bibnamefont
  {Clark}}\ and\ \bibinfo {author} {\bibfnamefont {N.~J.}\ \bibnamefont
  {Fisch}},\ }\href@noop {} {\bibfield  {journal} {\bibinfo  {journal} {Phys.
  Plasmas}\ }\textbf {\bibinfo {volume} {10}},\ \bibinfo {pages} {4848}
  (\bibinfo {year} {2003})}\BibitemShut {NoStop}%
\bibitem [{\citenamefont {Johnson}\ \emph {et~al.}(2017)\citenamefont
  {Johnson}, \citenamefont {Gordon}, \citenamefont {Palastro},\ and\
  \citenamefont {Hafizi}}]{Johnson17}%
  \BibitemOpen
  \bibfield  {author} {\bibinfo {author} {\bibfnamefont {L.~A.}\ \bibnamefont
  {Johnson}}, \bibinfo {author} {\bibfnamefont {D.~F.}\ \bibnamefont {Gordon}},
  \bibinfo {author} {\bibfnamefont {J.~P.}\ \bibnamefont {Palastro}}, \ and\
  \bibinfo {author} {\bibfnamefont {B.}~\bibnamefont {Hafizi}},\ }\href@noop {}
  {\bibfield  {journal} {\bibinfo  {journal} {Phys. Plasma}\ }\textbf {\bibinfo
  {volume} {24}},\ \bibinfo {pages} {033107} (\bibinfo {year}
  {2017})}\BibitemShut {NoStop}%
\bibitem [{\citenamefont {Malkin}\ and\ \citenamefont
  {Fisch}(2009)}]{Malkin09}%
  \BibitemOpen
  \bibfield  {author} {\bibinfo {author} {\bibfnamefont {V.~M.}\ \bibnamefont
  {Malkin}}\ and\ \bibinfo {author} {\bibfnamefont {N.~J.}\ \bibnamefont
  {Fisch}},\ }\href@noop {} {\bibfield  {journal} {\bibinfo  {journal} {Phys.
  Plasma}\ }\textbf {\bibinfo {volume} {80}},\ \bibinfo {pages} {046409}
  (\bibinfo {year} {2009})}\BibitemShut {NoStop}%
\bibitem [{\citenamefont {Zhang}\ \emph {et~al.}(2012)\citenamefont {Zhang},
  \citenamefont {He}, \citenamefont {Sheng},\ and\ \citenamefont
  {Yu}}]{zhang12}%
  \BibitemOpen
  \bibfield  {author} {\bibinfo {author} {\bibfnamefont {Z.~M.}\ \bibnamefont
  {Zhang}}, \bibinfo {author} {\bibfnamefont {X.~T.}\ \bibnamefont {He}},
  \bibinfo {author} {\bibfnamefont {Z.~M.}\ \bibnamefont {Sheng}}, \ and\
  \bibinfo {author} {\bibfnamefont {M.~Y.}\ \bibnamefont {Yu}},\ }\href@noop {}
  {\bibfield  {journal} {\bibinfo  {journal} {Appl. Phys. Lett}\ }\textbf
  {\bibinfo {volume} {100}},\ \bibinfo {pages} {134103} (\bibinfo {year}
  {2012})}\BibitemShut {NoStop}%
\bibitem [{\citenamefont {Zhang}\ \emph {et~al.}(2014)\citenamefont {Zhang},
  \citenamefont {Zhang}, \citenamefont {Hong}, \citenamefont {Yu},
  \citenamefont {Teng}, \citenamefont {He},\ and\ \citenamefont
  {Y.~Q.~Gu1}}]{Zhang14}%
  \BibitemOpen
  \bibfield  {author} {\bibinfo {author} {\bibfnamefont {Z.~M.}\ \bibnamefont
  {Zhang}}, \bibinfo {author} {\bibfnamefont {B.}~\bibnamefont {Zhang}},
  \bibinfo {author} {\bibfnamefont {W.}~\bibnamefont {Hong}}, \bibinfo {author}
  {\bibfnamefont {M.~Y.}\ \bibnamefont {Yu}}, \bibinfo {author} {\bibfnamefont
  {J.}~\bibnamefont {Teng}}, \bibinfo {author} {\bibfnamefont {S.~K.}\
  \bibnamefont {He}}, \ and\ \bibinfo {author} {\bibfnamefont {a.}~\bibnamefont
  {Y.~Q.~Gu1}},\ }\href@noop {} {\bibfield  {journal} {\bibinfo  {journal}
  {Phys. Plasmas}\ }\textbf {\bibinfo {volume} {21}},\ \bibinfo {pages}
  {123109} (\bibinfo {year} {2014})}\BibitemShut {NoStop}%
\bibitem [{\citenamefont {Zhang}\ \emph {et~al.}(2017)\citenamefont {Zhang},
  \citenamefont {Zhang}, \citenamefont {Hong}, \citenamefont {Deng},
  \citenamefont {Teng}, \citenamefont {He}, \citenamefont {Zhou},\ and\
  \citenamefont {Gu}}]{zhang17}%
  \BibitemOpen
  \bibfield  {author} {\bibinfo {author} {\bibfnamefont {Z.~M.}\ \bibnamefont
  {Zhang}}, \bibinfo {author} {\bibfnamefont {B.}~\bibnamefont {Zhang}},
  \bibinfo {author} {\bibfnamefont {W.}~\bibnamefont {Hong}}, \bibinfo {author}
  {\bibfnamefont {Z.~G.}\ \bibnamefont {Deng}}, \bibinfo {author}
  {\bibfnamefont {J.}~\bibnamefont {Teng}}, \bibinfo {author} {\bibfnamefont
  {S.~K.}\ \bibnamefont {He}}, \bibinfo {author} {\bibfnamefont {W.~M.}\
  \bibnamefont {Zhou}}, \ and\ \bibinfo {author} {\bibfnamefont {Y.~Q.}\
  \bibnamefont {Gu}},\ }\href@noop {} {\bibfield  {journal} {\bibinfo
  {journal} {Phys. Plasmas}\ }\textbf {\bibinfo {volume} {23}},\ \bibinfo
  {pages} {113104} (\bibinfo {year} {2017})}\BibitemShut {NoStop}%
\bibitem [{\citenamefont {Vieux}\ \emph {et~al.}(2017)\citenamefont {Vieux},
  \citenamefont {Cipiccia}, \citenamefont {Grant}, \citenamefont {Lemos},
  \citenamefont {Grant}, \citenamefont {Ciocarlan}, \citenamefont {Ersfeld1},
  \citenamefont {Hur}, \citenamefont {Lepipas}, \citenamefont {Manahan},
  \citenamefont {Raj}, \citenamefont {Gil}, \citenamefont {Subiel},
  \citenamefont {Welsh}, \citenamefont {Wiggins}, \citenamefont {Yoffe},
  \citenamefont {Farmer}, \citenamefont {Aniculaesei}, \citenamefont
  {Brunetti1},\ and\ \citenamefont {Yang}}]{Vieux17}%
  \BibitemOpen
  \bibfield  {author} {\bibinfo {author} {\bibfnamefont {G.}~\bibnamefont
  {Vieux}}, \bibinfo {author} {\bibfnamefont {S.}~\bibnamefont {Cipiccia}},
  \bibinfo {author} {\bibfnamefont {D.~W.}\ \bibnamefont {Grant}}, \bibinfo
  {author} {\bibfnamefont {N.}~\bibnamefont {Lemos}}, \bibinfo {author}
  {\bibfnamefont {P.}~\bibnamefont {Grant}}, \bibinfo {author} {\bibfnamefont
  {C.}~\bibnamefont {Ciocarlan}}, \bibinfo {author} {\bibfnamefont
  {B.}~\bibnamefont {Ersfeld1}}, \bibinfo {author} {\bibfnamefont {M.~S.}\
  \bibnamefont {Hur}}, \bibinfo {author} {\bibfnamefont {P.}~\bibnamefont
  {Lepipas}}, \bibinfo {author} {\bibfnamefont {G.~G.}\ \bibnamefont
  {Manahan}}, \bibinfo {author} {\bibfnamefont {G.}~\bibnamefont {Raj}},
  \bibinfo {author} {\bibfnamefont {D.~R.}\ \bibnamefont {Gil}}, \bibinfo
  {author} {\bibfnamefont {A.}~\bibnamefont {Subiel}}, \bibinfo {author}
  {\bibfnamefont {G.~H.}\ \bibnamefont {Welsh}}, \bibinfo {author}
  {\bibfnamefont {S.~M.}\ \bibnamefont {Wiggins}}, \bibinfo {author}
  {\bibfnamefont {S.~R.}\ \bibnamefont {Yoffe}}, \bibinfo {author}
  {\bibfnamefont {J.~P.}\ \bibnamefont {Farmer}}, \bibinfo {author}
  {\bibfnamefont {C.}~\bibnamefont {Aniculaesei}}, \bibinfo {author}
  {\bibfnamefont {E.}~\bibnamefont {Brunetti1}}, \ and\ \bibinfo {author}
  {\bibfnamefont {X.}~\bibnamefont {Yang}},\ }\href@noop {} {\bibfield
  {journal} {\bibinfo  {journal} {Scientific report}\ }\textbf {\bibinfo
  {volume} {7}},\ \bibinfo {pages} {2399} (\bibinfo {year} {2017})}\BibitemShut
  {NoStop}%
\bibitem [{\citenamefont {Ping}\ \emph {et~al.}(2009)\citenamefont {Ping},
  \citenamefont {Kirkwood}, \citenamefont {Wang}, \citenamefont {Clark},
  \citenamefont {Wilks}, \citenamefont {Wilks}, \citenamefont {Meezan},
  \citenamefont {Berger}, \citenamefont {Wurtele}, \citenamefont {Fisch},
  \citenamefont {Malkin}, \citenamefont {Valeo}, \citenamefont {Martins},\ and\
  \citenamefont {Joshi}}]{Ping091}%
  \BibitemOpen
  \bibfield  {author} {\bibinfo {author} {\bibfnamefont {Y.}~\bibnamefont
  {Ping}}, \bibinfo {author} {\bibfnamefont {R.~K.}\ \bibnamefont {Kirkwood}},
  \bibinfo {author} {\bibfnamefont {T.-L.}\ \bibnamefont {Wang}}, \bibinfo
  {author} {\bibfnamefont {D.~S.}\ \bibnamefont {Clark}}, \bibinfo {author}
  {\bibfnamefont {S.~C.}\ \bibnamefont {Wilks}}, \bibinfo {author}
  {\bibfnamefont {S.~C.}\ \bibnamefont {Wilks}}, \bibinfo {author}
  {\bibfnamefont {N.}~\bibnamefont {Meezan}}, \bibinfo {author} {\bibfnamefont
  {R.~L.}\ \bibnamefont {Berger}}, \bibinfo {author} {\bibfnamefont
  {J.}~\bibnamefont {Wurtele}}, \bibinfo {author} {\bibfnamefont {N.~J.}\
  \bibnamefont {Fisch}}, \bibinfo {author} {\bibfnamefont {V.~M.}\ \bibnamefont
  {Malkin}}, \bibinfo {author} {\bibfnamefont {E.~J.}\ \bibnamefont {Valeo}},
  \bibinfo {author} {\bibfnamefont {S.~F.}\ \bibnamefont {Martins}}, \ and\
  \bibinfo {author} {\bibfnamefont {C.}~\bibnamefont {Joshi}},\ }\href@noop {}
  {\bibfield  {journal} {\bibinfo  {journal} {Phys. Plasmas}\ }\textbf
  {\bibinfo {volume} {16}},\ \bibinfo {pages} {123113} (\bibinfo {year}
  {2009})}\BibitemShut {NoStop}%
\end{thebibliography}%

\end{document}